# SIMULATION STUDY OF ENERGY CHIRP INDUCED EFFECTS IN LASER-WAKEFIELD-ACCELERATOR-DRIVEN FREE ELECTRON LASERS

Shan-You Teng[1,2], Wai-Keung Lau[2], Shih-Hung Chen[1*], Wei-Yuan Chiang[2†]

[1]Department of Physics, National Central University, Taoyuan 320317, Taiwan

[2]National Synchrotron Radiation Research Center, Hsinchu 300092, Taiwan

*Abstract*

Beam energy compression via chicane magnets has been proved to be an effective method to reduce the slice energy spread of electron beams generated by laser wakefield accelerators (LWFAs). This technique has been widely adopted by leading research teams in experiments targeting future compact, high-gain free electron lasers (FELs). However, after energy compression, a strong beam energy chirp is introduced into the electron beam, which substantially hinders the microbunching process and impairs spectral coherence. Here, we present a detailed, unaveraged three-dimensional simulation that examines the effects of this energy chirp, and the results can be applied to the design of a proposed LWFA-driven VUV FEL. The energy chirp in a LWFA-produced electron beam causes FEL interactions at multiple resonant frequencies across the entire electron bunch, simultaneously, which prevents sustained radiation power growth at the designed frequency along the undulator. Consequently, spectral purity is significantly degraded. Additionally, due to undulator dispersion, the energy chirp leads to an elongation of the bunch length, which increases microbunch separation. This results in a noticeable redshift in the radiation frequency and further disruption of spectral purity. These effects are compared to the ideal scenario in which the energy chirp is removed following energy compression. Simulation results indicate that the implementation of a beam dechirper is a crucial step for improving the saturation of radiation power. Insights gained from this simulation of energy chirp-induced mechanisms will aid in the development of more effective compensation strategies, ultimately optimizing LWFA-driven FEL designs.

## I. INTRODUCTION

High-gain free electron lasers (FELs) are accelerator-based light sources capable of generating coherent radiation across a broad range of wavelengths, from THz to hard x-rays. The physical mechanism behind high-gain FEL is based on the subtle interaction between the laser field and an electron beam in an undulator under resonant conditions. The radiation power gain along the undulator depends on input beam qualities, including peak current, transverse emittance, energy spread, and the spatial coupling strength between the laser and electron beams. Generally speaking, high gain FEL requires slice beam transverse emittance to be smaller than the radiation wavelengths, and relative slice energy spread less than the FEL parameter [1]. Existing FEL facilities utilize electron beams with energies ranging from a few hundred MeV to tens of GeV to produce radiation from UV to x-ray [2-8]. However, the rf-based high energy linac systems are limited by accelerating gradients on the order of few tens MV/m and relatively large footprints (ranging from tens to hundreds of meters), resulting in high construction costs. In recent years, laser wakefield accelerators (LWFAs) have garnered significant attention due to their capability to produce high-brightness GeV-class electron beams in 6D phase space from extremely high accelerating gradient (i.e., GV/m) in plasma waves. These capabilities make LWFAs promising candidates as compact driver accelerators for single-pass short-wavelength high-gain FELs. Electron beams from LWFAs are characterized by sub-micron transverse emittance, femtosecond duration, and beam charge of a few tens picocoulombs. However, challenges remain in establishing beamlines for transporting and manipulating electron beams with milliradian divergence angle and percent-level relative energy spread for high-gain FEL applications [9-10]. Moreover, these beams are often subject to significant fluctuations due to laser instabilities, plasma density fluctuations, injection instabilities and nonlinear effects etc. to hinder stable operation of FELs [11].

[*]chensh@ncu.edu.tw  
[†]chiang.wy@nsrrc.org.tw

Despite fluctuations in beam parameters, some techniques have been developed to reduce slice energy spread. One promising approach is to tailor the longitudinal plasma density profile in the LWFA. By employing a controlled density down-ramp injection scheme, electron beams with reduced energy chirps can be generated, resulting in lower intrinsic energy spreads. Experiments at the Shanghai Institute of Optics and Fine Mechanics (SIOM) successfully produced electron bunches with charges ranging from 8.5-23.6 pC, energies 780–840 MeV, RMS energy spreads of 0.2–0.4%, and RMS divergences of 0.1–0.4 mrad [12]. This achievement marked a significant breakthrough in LWFA-driven FEL development at SIOM, where radiation from a self-amplified spontaneous emission (SASE) FEL was successfully operated at a wavelength of 27 nm [13]. Another method for reducing the RMS slice energy spread in LWFA beams is beam energy compression (sometimes referred to as bunch length decompression). This technique, typically through a magnetic chicane, introduces a longitudinal momentum dispersion to reduce the slice energy spread at the expense of the increased bunch length and reduced peak current. Slice energy spread can be reduced to a value as low as ~$10^{-3}$ for high gain FEL applications [9-10]. Considering the phase slippage of electrons with respect to the radiation field, the bunch length of a beam should be longer than the total slippage length throughout its propagation in the undulator so that the number of microbunches are maximized within the coherent length. Since the FEL parameter scales with the cubic root of peak current, the performance degradation of a high-gain FEL due to current reduction is relatively minor. While it is still beneficial to have an energy compressor that supports FEL operation at longer pulse duration, a large energy chirp (i.e. correlated energy spread) is inevitably remained after energy compression. Recently, a plasma wakefield dechirper has been proposed to remove the correlated energy spread, compensating for the residual chirp after energy compression [14]. In addition to mitigating correlated energy spread, the dechirper also reduces beam energy jitter. When combining with the energy compression, seeded FELs have been explored to enhance spectral purity and temporal coherence beyond SASE. One notable example is the COXINEL project at SOLEIL, which demonstrated a fully coherent FEL at 270 nm, driven by a LWFA with a mean beam energy ranging from 180 to 400 MeV and direct seeding [15]. It is worth noting that a noticeable redshift in radiation wavelength has been observed in the COXINEL FEL experiment, attributed to undulator-dispersion induced bunch lengthening (or modulation period stretching, as termed in Ref. 14), caused by the large correlated energy spread left after energy compression. It is worth to point out that dechirper has not been installed in their setup. In addition to the discussed energy spread reduction techniques, transverse gradient undulators (TGUs) have been proposed to mitigate energy spread effects by spreading electron orbits in transverse direction according to electron energies and gradually varying the undulator field in the same direction This allows electrons with slightly different energies to remain in phase with the radiation field [16]. However, the use of TGUs in FELs is beyond the scope of this study, which focuses on LWFA-driven FEL systems incorporating energy compressors.

In this study, we investigate the microbunch dynamics of an energy-compressed beam in the undulator and its impact on the spectral properties of a SASE FEL, using the design of a compact VUV FEL proposed at National Central University (NCU). The proposed FEL is a 70-nm SASE driven by an energy compressed 250 MeV LWFA beam. When such an energy-chirped beam propagates through an undulator, which inherently exhibits longitudinal dispersion, significant elongation of the bunch length and increased microbunch separation are expected. These effects result in a reduction of the peak current, a red shift in the radiation wavelength, and a broadening of the emission spectrum. We also anticipate improvements in peak power, pulse energy and the spectral bandwidth of the FEL radiation with the activation of a beam dechirper. Transportation of the ultra-short bunch with high electron density from the LWFA is carried out by using the full 3D space charge code – IMPACT [17-18], which accounts for the effects of large beam energy spread and 1D coherent synchrotron radiation (CSR). Simulation of the FELs are performed using PUFFIN [19-20], an unaveraged 3D code that provides deeper insight into FEL performance for ultrashort beams with large energy spreads delivered by the LWFA [21-22].

Section 2 provides a general description of the energy-chirp-induced effects in FEL for a LWFA beam after energy compression. In Sec. 3, the proposed NCU FEL beamline and design parameters for the SASE FEL are discussed. Section 4 presents simulation results of microbunch behavior and FEL performance for an energy-chirped beam, as well as the case where the chirp is ideally removed. Section 5 concludes the study.

## II. ENERGY CHIRP INDUCED EFFECTS

This section presents a basic analysis of the dispersive effects experienced by an electron beam after energy compression and subsequent transport through a planar undulator. Energy compression is typically achieved through the use of a magnetic chicane. Transformation of the coordinates of an electron in longitudinal phase space through a chicane with $R_{56_{chicane}}$ is described by the following mapping equations:

$$z_f = z_i + R_{56_{chicane}} \delta_i, \\ \delta_f = \delta_i - h z_f, \qquad (1)$$

where $z_i$ and $\delta_i$ are the initial longitudinal position and relative energy deviation of an electron. $z_f$, $\delta_f$ are the corresponding values after the chicane. Then, the slice energy spread $\sigma_{\delta_f}$ of the energy compressed beam can be approximately expressed as:

$$\sigma_{\delta_f} \approx \sqrt{\left((1 - hR_{56_{chicane}})^2 \sigma_{\delta_i}^2 + h^2 \sigma_{z_i}^2\right)}, \qquad (2)$$

where $\sigma_{\delta_i}$ is the initial relative energy spread and h is the linear energy chirp introduced by the energy compressor. After the energy compressor, the LWFA-generated electron beam exhibits a significant reduction in slice energy spread, to meet the criteria of FEL lasing. However, the bunch length simultaneously elongates according to Eq.3, i.e.,

$$\sigma_{z_f} \approx \sqrt{\left(\sigma_{z_i}^2 + \left(R_{56_{chicane}} * \sigma_{\delta_i}\right)^2\right)}, \qquad (3)$$

where $\sigma_{z_i}$ is the initial bunch length and $\sigma_{z_f}$ denotes final bunch length. This implies that reducing the slice energy spread requires increasing the bunch length. However, the slice energy spread cannot be minimized indefinitely without sacrificing peak current. Since a high peak current is essential for efficient FEL operation, this trade-off places a practical limit on energy compression. As a result, a residual correlated energy chirp may persist in the beam after partial energy compression. For LWFA beams with ultrashort bunch lengths, the energy chirp can be approximated as:

$$h \approx \frac{1}{R_{56_{chicane}}}. \tag{4}$$

To maintain a bunch length on the order of micrometers in LWFA-based beams, the required longitudinal dispersion $R_{56_{chicane}}$ is typically on the order of $10^{-4}$m, corresponding to a strong linear energy chirp approaching GeV/m. Such a strong energy chirp has a significant impact on the longitudinal beam dynamics in FELs.

The corresponding undulator dispersion follows directly from the longitudinal velocity of electrons in a planar undulator. In the relativistic regime, transverse oscillations lead to a small reduction in the longitudinal velocity, given by:

$$v_z = c\left(1 - \frac{1}{2\gamma^2}\left(1 + \frac{K^2}{2}\right)\right), \tag{5}$$

where c is the speed of light, $\gamma$ the relativistic Lorentz factor and the undulator parameter K is defined as:

$$K = \frac{eB_u\lambda_u}{2\pi m_e c}, \tag{6}$$

where $B_u$ denotes the peak magnetic field amplitude, e represents the electron charge, $m_e$ the electron mass, and $\lambda_u$ being the undulator period. Expanding $v_z$ about $\gamma$ and evaluating the transit time over $N_u$ undulator periods, the following expression for the longitudinal dispersion:

$$\frac{dz}{d\delta} \approx \frac{\lambda_u N_u}{\gamma^2}\left(1 + \frac{K^2}{2}\right). \tag{7}$$

Finally, for an ultrashort electron beam perform energy compression in a four-dipole chicane with longitudinal dispersion $R_{56_{chicane}}$, the resulting elongation of microbunch separation $\Delta z$ after traversing the undulator is given by:

$$\Delta z \approx \frac{\lambda_u N_u}{\gamma^2}\left(1 + \frac{K^2}{2}\right)\lambda_r/R_{56_{chicane}}, \tag{8}$$

where $\lambda_r$ is the resonant wavelength. As the chirped electron beam propagates through the undulator, energy-dependent longitudinal dispersion leads to bunch lengthening and increases separation between microbunches. This spatial spreading degrades FEL performance in several ways. First, variation of the resonant condition along the beam significantly limits the coherent interaction length Second, the enlarged microbunch spacing shifts the resonant wavelength to longer values, producing a red shift in the output spectrum. Lastly, bunch elongation lowers the peak current, thereby weakening the FEL gain. These effects make the implementation of a beam dechirper and other compensation strategies in LWFA-driven FELs essential.

## III.  LWFA DRIVEN SASE FEL—AN ILLUSTRATIVE EXAMPLE

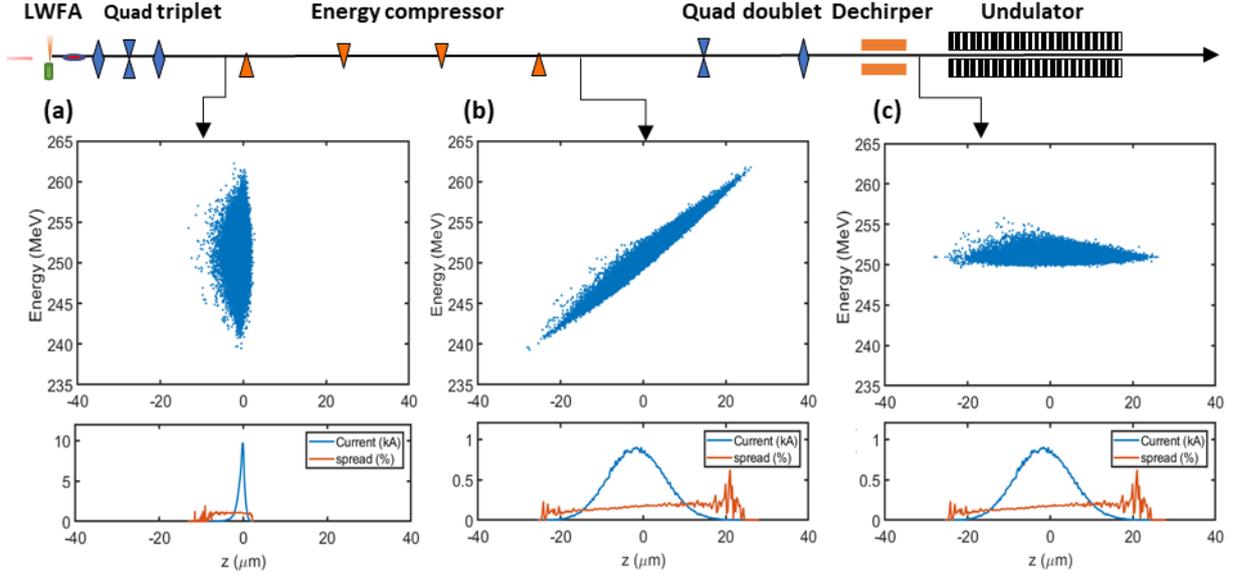

FIG. 1. Schematic layout of the proposed beamline and the corresponding electron beam properties at each stage. (a) A focused ultrashort electron beam after the quadrupole triplet in the transport line from the LWFA to the undulator entrance. (b) An energy-compressed electron beam with reduced slice energy spread after passing through a four-dipole chicane. (c) An electron beam with its energy chirp removed by an ideal dechirper. The simulations were performed using IMPACT.

The schematic layout of NCU beamline is shown in Fig.1. It is a VUV SASE FEL driven by LWFA electron beam undergoing successive stages of beam manipulations. As shown in Fig. 1(a), the LWFA can generate intense electron beam with peak currents up to 10kA. However, the relatively large energy spread of LWFA-generated beams poses a major challenge for FEL applications, as it reduces gain and limits coherent amplification. For efficient FEL operation, the energy spread must be smaller than the Pierce parameter. To address this issue, a 4-dipole energy compressor can be employed to reshape the longitudinal phase space of the electron beam. Figure 1(b) depicts the phase space distribution of the energy compressed beam after the 4-dipole chicane. However, the energy compressor introduces a linear chirp of 500MV/mm, which is difficult to compensate using conventional RF linearizers or dielectric dechirpers. A promising alternative is an active plasma dechirper driven by an intense laser, capable of providing a dechirping strength of up to 62GeV/mm/m [23-24], sufficient to fully compensate the chirp within 5mm. Figure 1(c) illustrates the electron distribution after an ideal dechirper, with other key beam parameters remaining unchanged.

A. Simulation of Beam Transport and Manipulation

Within our beamline simulation framework, the IMPACT program serves as the primary simulation tool. The key parameters of the electron beam used in this study are summarized in Table 1.

Table 1. Simulation parameters of the NCU VUV FEL

|  | Parameter | Value |
|---|---|---|
| **Electron beam** | Energy | 250 MeV |
|  | Bunch charge | 20-50 pC |
|  | Bunch duration | 2.12 fs |
|  | Normalized emittance | 0.5 mm mrad |
|  | Initial beam size | 0.72 μm |
|  | Initial beam divergence | 1.4 mrad |
|  | Energy spread | 0.1-1 % |
| **Undulator** | Period length $\lambda_u$ | 20 mm |
|  | Number of periods $N_u$ | 350 |
|  | Undulator parameter $K_u$ | 1.17 |

The initial divergence is estimated to be 1.4 mrad, which can significantly increase the beam size in the early stage of beam transport line. To counteract this effect, a strong permanent quadrupole triplet with a maximum gradient of 250 T/m is placed 5 cm downstream of the LWFA gas jet to rapidly focus the electron beam. Figure 2 shows the optimized beam size and emittance evolution along the beam transport line and undulator. Due to high intrinsic divergence and emittance growth from dispersion, the beam size can expand to several hundred μm. Matching of betatron motions in both planes is achieved by analyzing the evolution of beam sizes within the undulator. The planar undulator provides vertical focusing, primarily reducing the electron beam size in the y-direction. By increasing the horizontal focusing strength during beam transport, the alignment of the beam waist at the same position is achievable within the undulator. This approach maximizes electron density and enhances FEL performance.

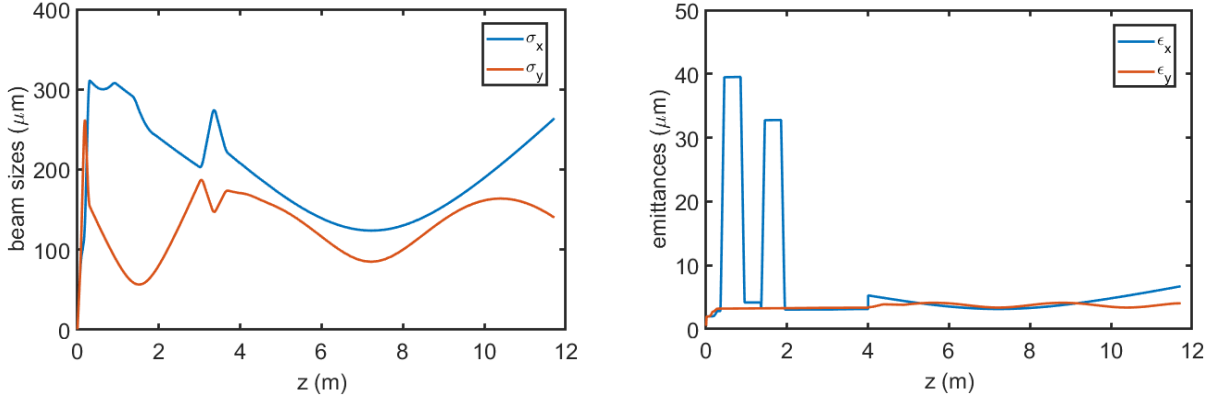

FIG. 2. Simulated evolution of transverse beam parameters along the optimized LWFA beam transport line, preserving beam quality for FEL amplification.

## B.  FEL simulation

Since the LWFA beam for the VUV FEL is only a few wavelengths long, its parameters vary rapidly during FEL interactions, leading to a broader spectral content. Therefore, the broadband FEL code PUFFIN is more suitable in the study than conventional codes based on the slowly varying envelope approximation. PUFFIN code enables direct simulation of detailed particle dynamics in FEL without the need of phase-space slicing. It fully accounts for the complete radiation field along the undulator using a PIC-like algorithm, making it particularly well-suited for modeling FEL systems with ultrashort electron bunches and large energy spreads. In the simulation, the electron beam generated by the IMPACT code is used as the input for PUFFIN. To analyze the beam dynamics and FEL amplification process, three electron beam configurations corresponding to Fig. 1(a) - 1(c) are investigated.

The peak power growth of for the FEL without using energy compressor and dechiper at various initial beam energy spread (Fig.1(a)) is shown in Fig. 3. The radiation powers rise sharply within the first meter of the undulator, followed by a linear growth that depends on the initial beam energy spread, reaching saturation at approximately 3 to 4 meters. We attribute this rapid initial growth to coherent undulator radiation (CUR) generated by the short LWFA beam. The CUR power at the entrance of the undulator is estimated by the following formula [25], i.e.,

$$P_{CHG} = \frac{Z_0 I_0^2 K^2 b_n^2 F_{B2}^2 L_u^2}{32\pi \sigma_T^2 \gamma^2}, \qquad (9)$$

where $Z_0 = 377\Omega$ is the free space wave impedance, $I_0$ is the peak current of electron beam, $K$ is the undulator parameter, $b_n$ the nth harmonic bunching factor, and $F_{B2}, L_u, \sigma_T$ and $\gamma$ are the Bessel factor, length of undulator, transverse beam size and the reference energy, respectively. For the estimation, the bunching factor and electron beam sizes are taken from PUFFIN simulation data at a diagnostic point located 0.5 m downstream of the undulator entrance. Using these values, the estimated CUR power is around 5MW. In Fig. 3, the CUR power rises to ~ 4 MW within the first meter of the undulator, in good agreement with the analytical estimate. With an energy spread of 0.5%, FEL amplification still occurs, reaching saturation at around 3 m. When the energy spread exceeds the FEL Pierce parameter ($\rho = 0.0082$), no gain is observed.

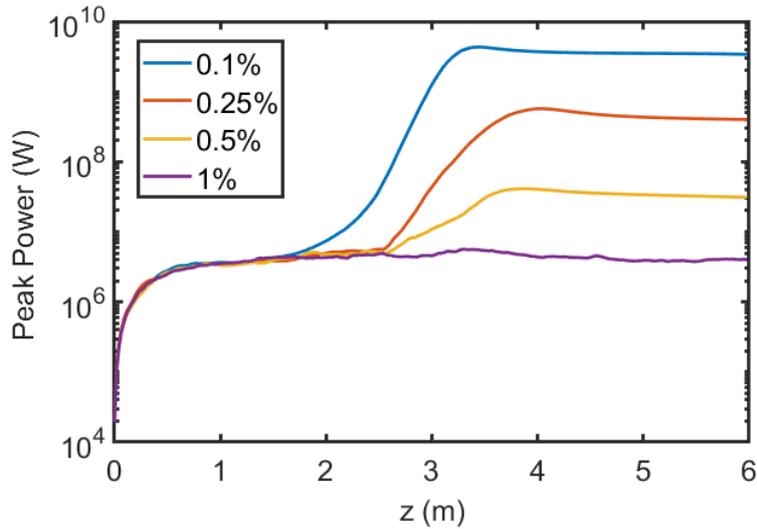

FIG. 3. Evolution of the SASE FEL peak power for electron beams with initial electron beam energy spreads of 0.1%, 0.25%, 0.5%, and 1%, respectively.

## IV. ENERGY COMPRESSION AND DECHIRPING

To evaluate the practical impact of energy compression on FEL performance, we perform simulations using the energy compressed electron beam shown in Fig. 1(b). After passing through a four-dipole energy compressor, while maintaining a peak current of approximately 1 kA, the slice energy spread is reduced from about 1% to 0.15%. The residual correlated energy chirp after energy compression is then removed by an ideal dechirper, yielding the dechirped beam as shown in Fig. 1(c). The corresponding FEL power evolution for both cases are presented in Fig. 4. The results show that reducing the slice energy spread significantly enhances FEL gain. In particular, the dechirped beam produces an output power more than an order of magnitude higher than that obtained without a dechirper. Beyond the difference in peak power, there are also marked changes in the longitudinal startup position and saturation length. Without the dechirper, linear growth begins earlier but saturates sooner, whereas with a dechirper, growth starts later but is sustained longer, leading to higher saturation power.

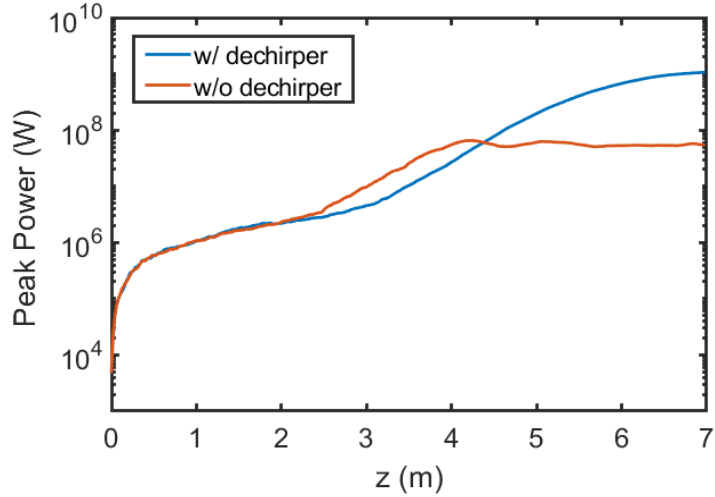

FIG. 4. Evolution of FEL peak power along the undulator for the energy compressed LWFA beam (orange curve) without dechirper and with dechirper (blue curve).

This phenomenon arises from the longitudinal variation of the resonant wavelength, as reflected in the output radiation spectra shown in Fig. 5. The uncorrected beam produces radiation with multiple spectral peaks and a pronounced redshift in the dominant component. Microbunching degrades when decoherence between radiation frequencies occurs, particularly when the resonant frequency spread caused by the beam energy chirp exceeds the FEL gain bandwidth. In contrast, the dechirped beam yields a single, narrow spectral peak centered at the resonant wavelength, in agreement with the theoretical value, demonstrating improved spectral coherence and a more efficient FEL gain process.

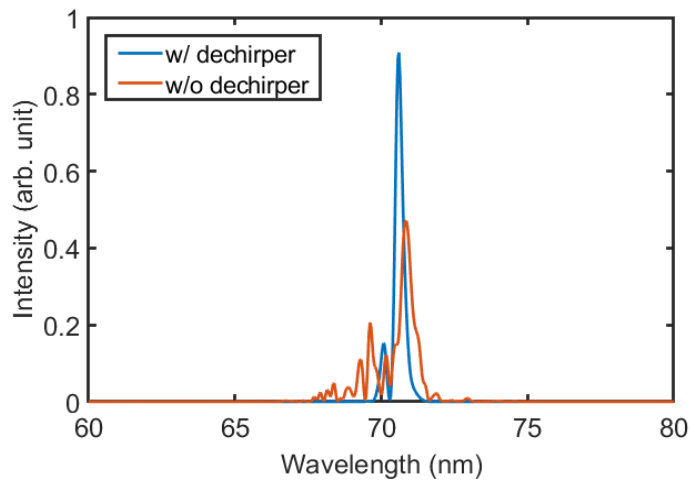

FIG. 5. Comparison of output spectra for the case without dechiper (orange curve) and with dechirper (blue curve).

The effect of a correlated energy chirp on FEL gain can be evaluated by tracking the evolution of the microbunch spectrum along the undulator, as shown in Figs. 6 and 7. In the energy-compressed case, the spectral peak shifts from approximately 69 nm to over 72 nm as the beam propagates. This shift reflects increased microbunch separation caused by undulator dispersion acting on a chirped beam, which can be quantified using Eq. (8). As the bunch stretches longitudinally, electrons fall out of phase with the radiation field, degrading microbunching coherence and limiting FEL amplification. The spectrum also broadens, indicating that multiple longitudinal modes simultaneously satisfy their respective resonance conditions and undergo FEL gain—consistent with the features observed in the output spectra.

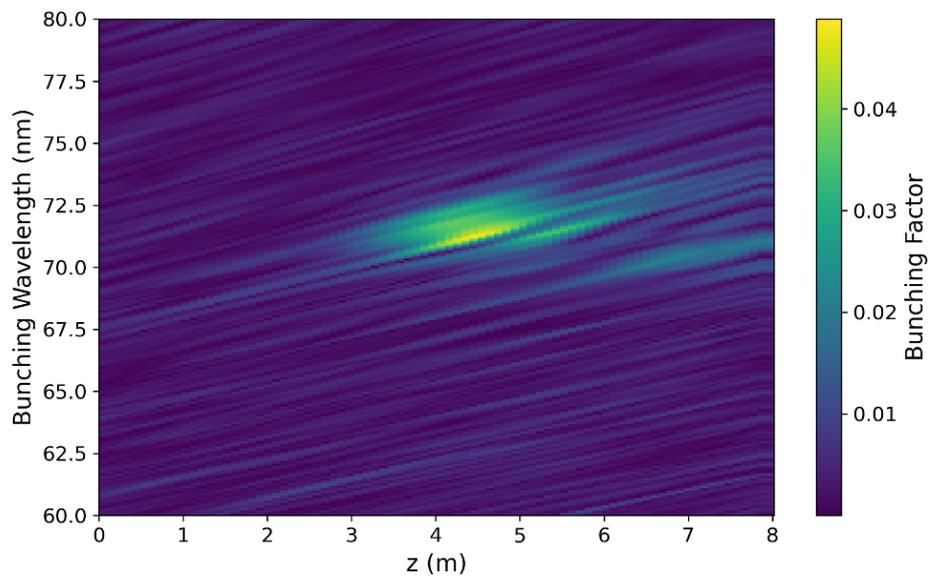

FIG. 6. Beam spectrum evolution along the undulator for the case without dechirper. The color scale indicates the bunching strength at each wavelength and undulator position.

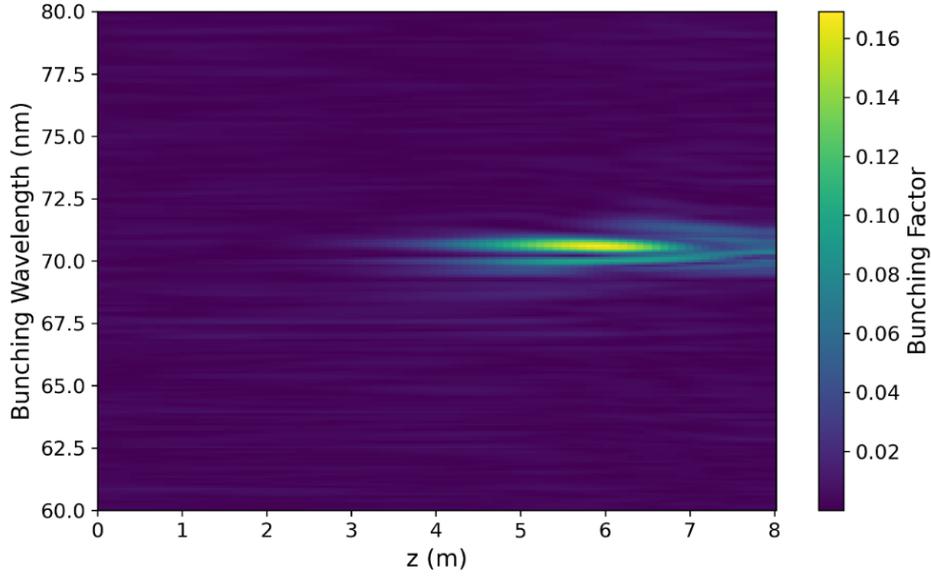

FIG. 7. Beam spectrum evolution along the undulator for the case with dechirper. The color scale indicates the bunching strength at each wavelength and undulator position.

In contrast, when a dechirper is applied after energy compression, the bunching wavelength remains nearly constant at the target resonant condition throughout the undulator. This preserves longitudinal phase-space alignment, allowing a larger fraction of electrons to undergo identical FEL interactions simultaneously. The result is stronger, spectrally stable bunching and sustained amplification with high temporal and spectral coherence.

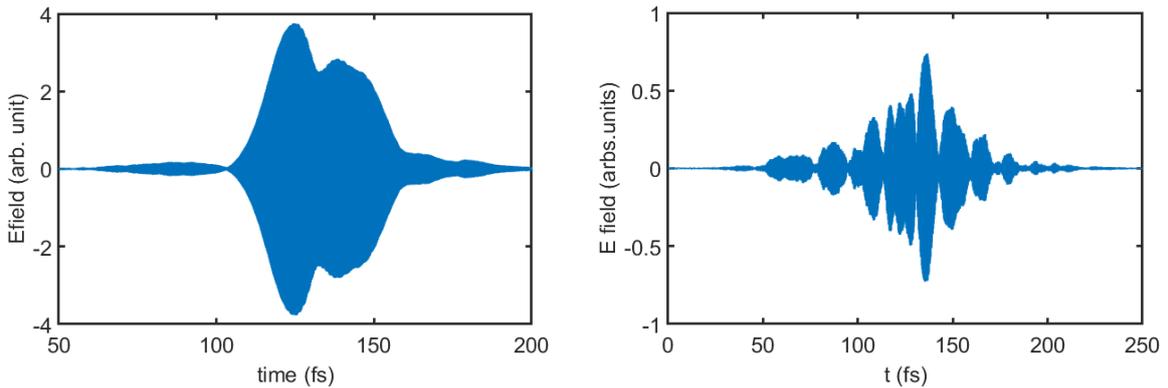

FIG. 8. Electric field emitted from the case with dechirper (left) and without dechirper (right).

Finally, the temporal structures of the radiation fields are shown in Fig. 8. In the case without a dechirper, the field profile exhibits multiple temporal envelopes with varying amplitudes. This arises from the energy chirp, which causes different slices of the bunch to radiate at distinct wavelengths and phases, thereby reducing the temporal coherence of the output pulse.

In contrast, the field from the dechirped case exhibits a well-defined envelope, indicating coherent emission from nearly the entire bunch. These time-domain results align with the spectral observations in Figure 5. The energy-compressed beam produces a broadened, multi-peaked spectrum, whereas the dechirped beam yields a narrowband, spectrally coherent output. Overall, the findings confirm that energy chirp in the electron beam markedly degrades both the spectral and temporal quality of FEL

## V. CONCLUSION

In summary, this study systematically investigates the impact of energy chirp on electron beam dynamics and the output performance of FELs driven by LWFAs. Using a combination of beam dynamics simulations and unaveraged 3D FEL calculations, we demonstrate that energy compression via magnetic chicanes effectively reduces the slice energy spread needed for high-gain FEL amplifiers. However, magnetic chicanes also introduce a strong correlated energy chirp within the electron bunch. This residual energy chirp causes the bunch to elongate and increases microbunch separation due to undulator dispersion. As a result, the output radiation exhibits a red shift in wavelength, broadening of the FEL spectrum, and significant degradation in both temporal and spectral coherence.

The implementation of a beam dechirper to remove this residual energy chirp proves crucial for optimizing FEL performance. Dechirping restores the phase-space alignment of the electron beam, allowing the FEL interaction to be maintained at the designed resonant frequency across the entire bunch. This, in turn, supports efficient microbunching and single-mode amplification. Simulation results confirm that, compared to the energy-compressed case without dechirping, the use of a dechirper leads to more than an order of magnitude increase in saturated FEL power, narrows the radiation spectrum, and improves both spectral purity and temporal coherence of the FEL output.

These findings highlight the necessity of understanding and compensating for energy chirp in order to achieve high-brightness, spectrally coherent FEL sources based on LWFA beams. The future development of compact FEL facilities can benefit from advanced dechirping techniques or other methods to mitigate the effects of strong energy chirp.

## ACKNOWLEDGEMENTS

The authors would like to express our appreciation to Dr. Ji Qiang of LBNL for his guidance to use IMPAC-T code for this study and Dr. Shao-Wei Chou for intensive discussions on the design study of a compact VUV FEL. This work was partially supported by the National Science and Technology

Council (NSTC), Taiwan, under Grant number NSTC 113-2112-M-213-024, NSTC 113-2119-M-001-007, and NSTC 113-2112-M-008-010.